\documentclass[12pt]{article}
\usepackage{epsfig}
\usepackage{hyperref}
\usepackage{amsmath}
\usepackage{bm}
\usepackage{textcomp}

\usepackage{epstopdf}
\def\beq{\begin{equation}}
\def\eeq#1{\label{#1}\end{equation}}
\def\eeqn{\end{equation}}

\def\beqa{\begin{eqnarray}}
\def\eeqa#1{\label{#1}\end{eqnarray}}
\def\eeqan{\end{eqnarray}}

\let\bar=\overbar

\def\Dslash{\not{\hbox{\kern-4pt $D$}}}
\def\dslash{\not{\hbox{\kern-2pt $\del$}}}

\def\msb{{\bar{\ssstyle M \kern -1pt S}}}

\def\Title#1{\begin{center} {\Large {\bf #1} } \end{center}}

\begin{document}
\Title{Measurements of B lifetimes at LHCb}
\bigskip\bigskip
\begin{raggedright}  
{\it Dordei Francesca, on behalf of the LHCb collaboration \index{Francesca, D.}\\
Heidelberg University, Physikalisches Institut\\
Im Neuenheimer Feld 226, 69120 Heidelberg, GERMANY}
\bigskip\bigskip
\end{raggedright}
{\vspace{-0.8cm}
\begin{abstract}
Measurements of the $B_s^0$ effective lifetime in decays to $CP$-odd and $CP$-even flavour specific final states allow to probe the width difference $\Delta \Gamma_s$ and the $CP$-violating phase $\phi_s$ of $B^0_s-\overline{B}^0_s$ mixing box-diagram. Measurements of the effective lifetime in the decay channels $B_s^0 \rightarrow K^+ K^-$ and $\overline{B}_s^0 \rightarrow J/\psi f_0(980)$ are presented, as well as a measurement of $\Delta \Gamma_s$ and $\Gamma_s$ performed by a tagged time-dependent angular analysis of $B_s^0 \rightarrow J/\psi \phi$ decays and a measurement of the sign of $\Delta \Gamma_s$, using data collected during 2011 with the LHCb detector.
\end{abstract}
}
\section{Introduction}
Precision measurements of $B$-hadron lifetimes are an important test of the theoretical approach to $B$-hadron observables, known as Heavy Quark Expansion (HQE). In this approach the decay rate is calculated as an expansion in inverse powers of the heavy b-quark mass~\cite{Lenz}: 
\begin{equation}
\label{eq:lifetime}
\Gamma = \Gamma_0 + (\Lambda^2/m_b^2)\Gamma_2 + (\Lambda^3/m_b^3)\Gamma_3 + ...
\end{equation}
where $\Gamma_0$ represents the decay of a free heavy b-quark, and according to this contribution all $B$-mesons have the same lifetime.
Different corrections at order $\Lambda^i/m^i_b$ with $i=2,3,...$ due to the kinetic and the chromomagnetic operators, spectator quark involved in the weak annihilation and Pauli interference alter the lifetime at approximately the $10\%$ level.
Furthermore measurements of the effective lifetimes in $B^0_s$-meson decays allow to probe the width difference $\Delta \Gamma_s$ and the $CP$-violating phase $\phi_s$ of $B^0_s-\overline{B}^0_s$ mixing box-diagram~\cite{Fleischer}. 
We can consider a $B^0_s (\overline{B}^0_s) \rightarrow f$ transition, with a final state $f$ into which both a $B^0_s$ and a $\overline{B}^0_s$-meson can decay. When the initial flavour of the $B^0_s$-meson is unknown the corresponding untagged rate can be written as follows:
\begin{equation}
\label{eq:decayrate}
  \Gamma (B^0_s(\overline{B}^0_s)(t) \rightarrow f) \propto (1-\mathcal{A}_{\Delta \Gamma_s})e^{-\Gamma_Lt}+(1+\mathcal{A}_{\Delta \Gamma_s})e^{-\Gamma_Ht}
\end{equation}
The quantities $\Gamma_H$ and $\Gamma_L$ are the decay widths of the heavy and light $B^0_s$ mass eigenstates and $\Delta \Gamma_s = \Gamma_L - \Gamma_H$ is the decay width difference. The parameter $\mathcal{A}_{\Delta \Gamma_s}$ is defined as $\mathcal{A}_{\Delta \Gamma_s} = -2\Re e(\lambda)/(1+|\lambda|^2)$, where $\lambda = (q/p)(\overline{A}/A)$. The complex coefficients $p$ and $q$ define the mass eigenstates of the $B^0_s-\overline{B}^0_s$ system in terms of the flavour eigenstates (see e.g. Ref.~\cite{LenzNierste}) and $A(\overline{A})$ is the amplitude for a $B^0_s(\overline{B}^0_s)$ to decay to $f$. This parameter $\mathcal{A}_{\Delta \Gamma_s}$ is a function of $\phi_s$~\cite{Fleischer}.
We define the effective lifetime of the decay $B^0_s \rightarrow f$ as the time expectation value of the untagged rate:
\begin{equation}
  \tau_f \equiv \frac{\int_0^\infty t \,\,\langle \Gamma (B^0_s(\overline{B}^0_s)(t) \rightarrow f) \rangle dt}{\int_0^\infty \langle \Gamma (B^0_s(\overline{B}^0_s)(t) \rightarrow f) \rangle dt}
\end{equation}
which is equivalent to the lifetime that results from fitting the untagged decay time distribution in Eq.~\ref{eq:decayrate} with a single exponential.
By making use of the usual definition $y_s = \Delta \Gamma_s / 2\Gamma_s$ and using $\tau_{B^0_s}= \Gamma_s^{-1}$, one can espress the effective lifetime as:
\begin{equation}
\tau_f = \tau_{B^0_s} + \tau_{B^0_s} \mathcal{A}_{\Delta \Gamma_s} y_s + \tau_{B^0_s} [2-(\mathcal{A}_{\Delta \Gamma_s})^2] y_s^2 + \mathcal{O}(y_s^3).
\end{equation}
So it is possible to use the corresponding lifetime measurements to constrain $y_s$ (or $\Delta \Gamma_s$) with respect to $\phi_s$, with the advantage that only an untagged analysis is needed.
\section{$B^0_s$ effective lifetime, $\Delta \Gamma_s$ and $\Gamma_s$ determination}
\begin{figure}[ht]
\begin{center}
\epsfig{file=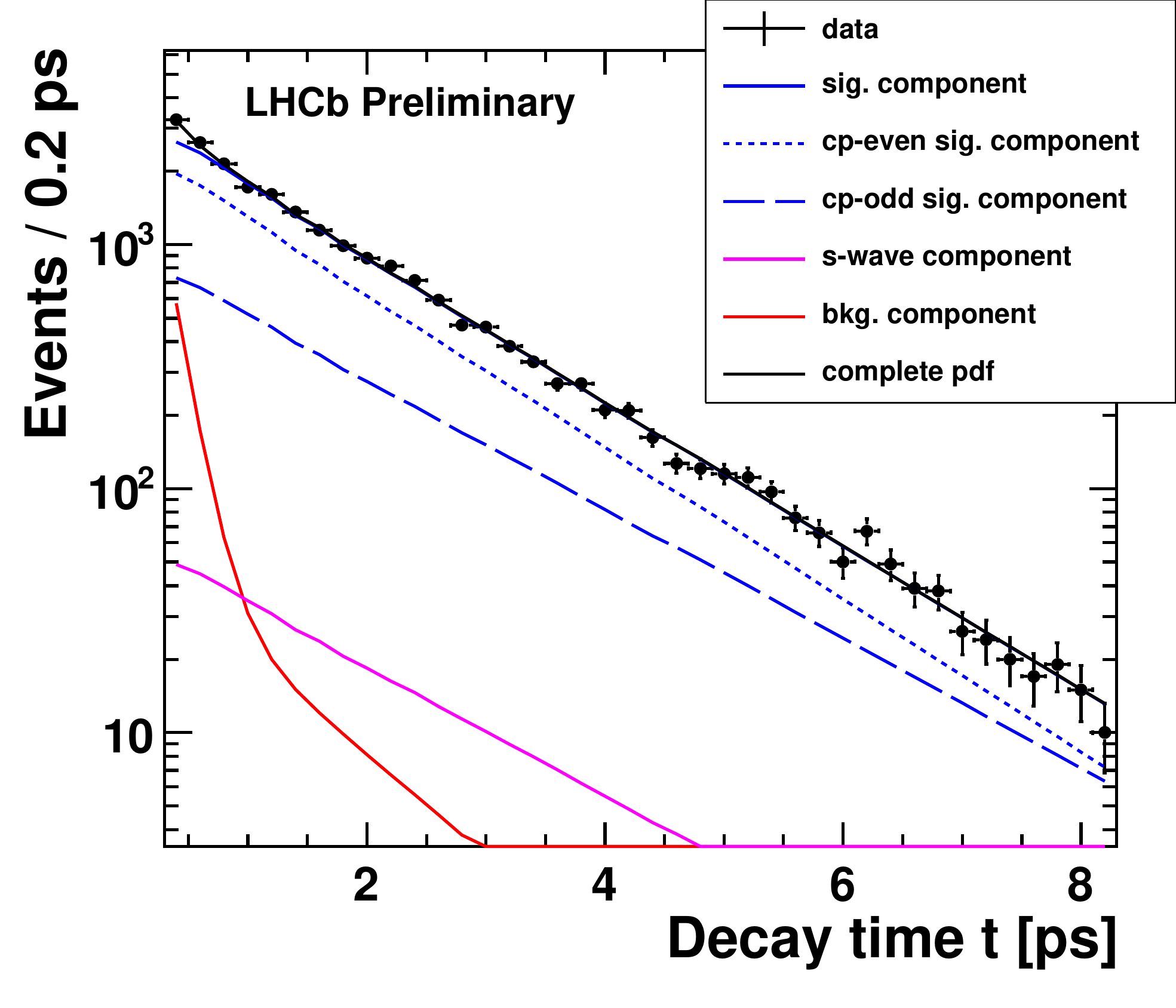,height=1.8in}
\epsfig{file=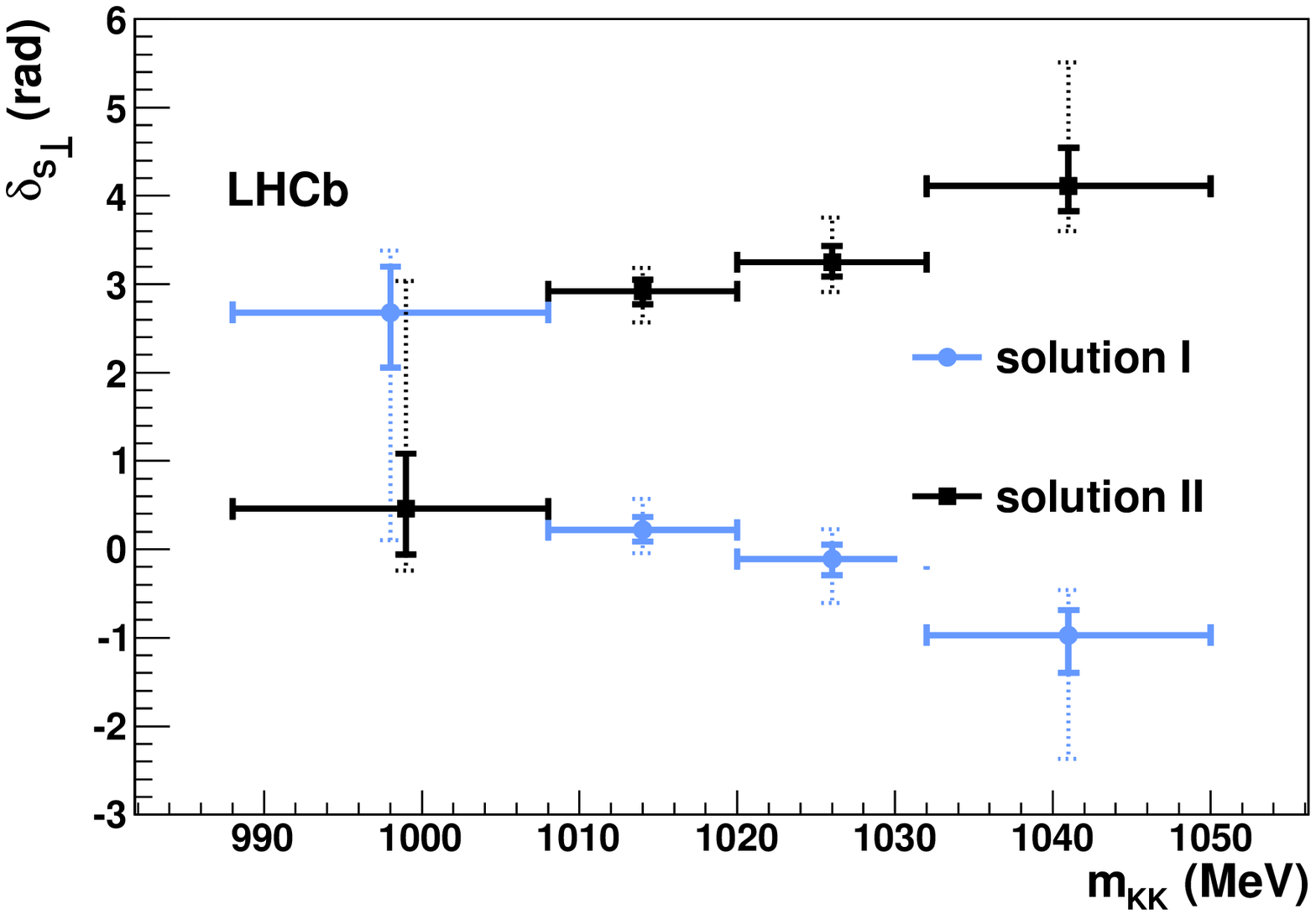,height=1.6in}
\caption{(a) Decay time distributions, overlayed with the one dimensional projections of the fitted PDF (black), split in signal part (blue) and background part (red).\newline (b) Measured phase differences between S-wave and perpendicular P-wave amplitudes in four intervals of $m_{K^+K^-}$ for solution I (positive $\Delta \Gamma_s$) and solution II (negative $\Delta \Gamma_s$).}
\label{fig:jpsiphi}
\end{center}
\end{figure}

The $B^0_s$ effective lifetime presented here has been measured through the decay channels \mbox{$B^0_s \rightarrow K^+ K^-$} and $\overline{B}^0_s \rightarrow J/\psi f_0(980)$ (charge conjugate modes are implied throughout), using $1.0$ $\mathrm{fb}^{-1}$ of data collected in $pp$ collisions at $\sqrt{s}=7$ TeV with the LHCb detector~\cite{LHCb}.
At LHCb, $B$-hadrons are produced with an average momentum of around $100$ $\mathrm{GeV}/\mathrm{c}$ and have decay vertices displaced from the primary interaction vertex (around 1-2 cm for $B^0_s$), whereas combinatorial background candidates are produced by the random combination of tracks, which tend to originate from the primary vertex.
So conventional selections exploit these features to select $B$-hadrons by requiring that their decay products are energetic and significantly displaced from the primary interaction point. 
However, this introduces a time-dependent acceptance which needs to be taken into account in the analysis.
Moreover, there could be also implicit biases, for example due to geometrical and reconstruction acceptances.
Experimentally it is challenging to correct for these biases. Here two strategies are presented to make an unbiased lifetime measurement: an approach where $B$-hadrons are selected without biasing the decay time distribution and a measurement of the $B^0_s$ lifetime relative to the well known $B^0$ lifetime.\\
A measurement of the effective lifetime of the $B^0_s \rightarrow K^+ K^-$ decay is of considerable interest as it is sensitive to new physics phenomena affecting the $B^0_s$ mixing phase and entering the decay at loop level. 
The $K^{+}K^{-}$ final state is $CP$-even and in the Standard Model $\mathcal{A}_{\Delta \Gamma_s}$ is predicted to be equal to $\mathcal{A}_{\Delta \Gamma_s} = -0.972 \pm 0.012$~\cite{KK}, hence the effective lifetime is approximately equal to $\Gamma^{-1}_L$. 
\begin{figure}[ht]
\begin{center}
\epsfig{file=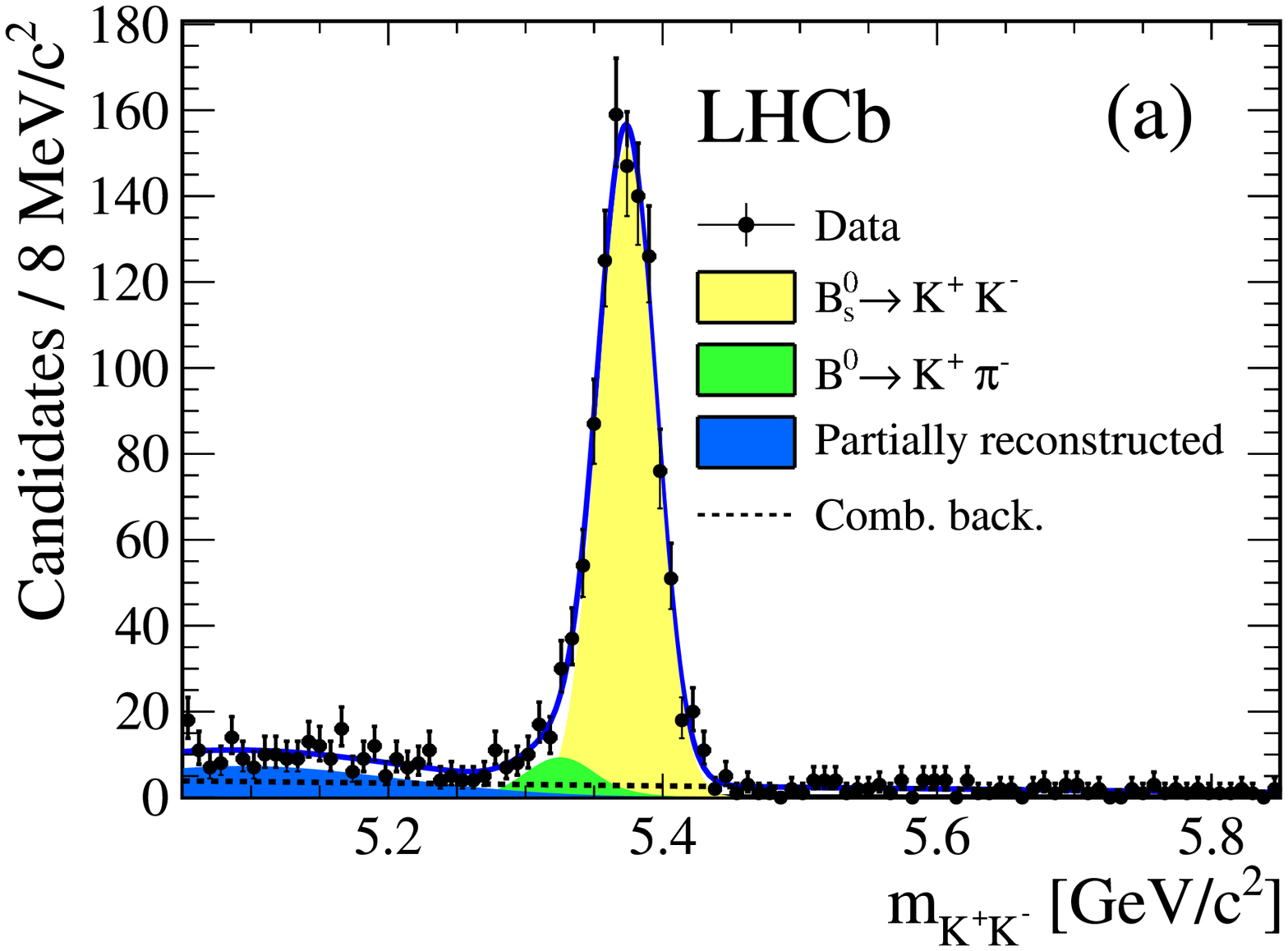,height=1.5in}
\epsfig{file=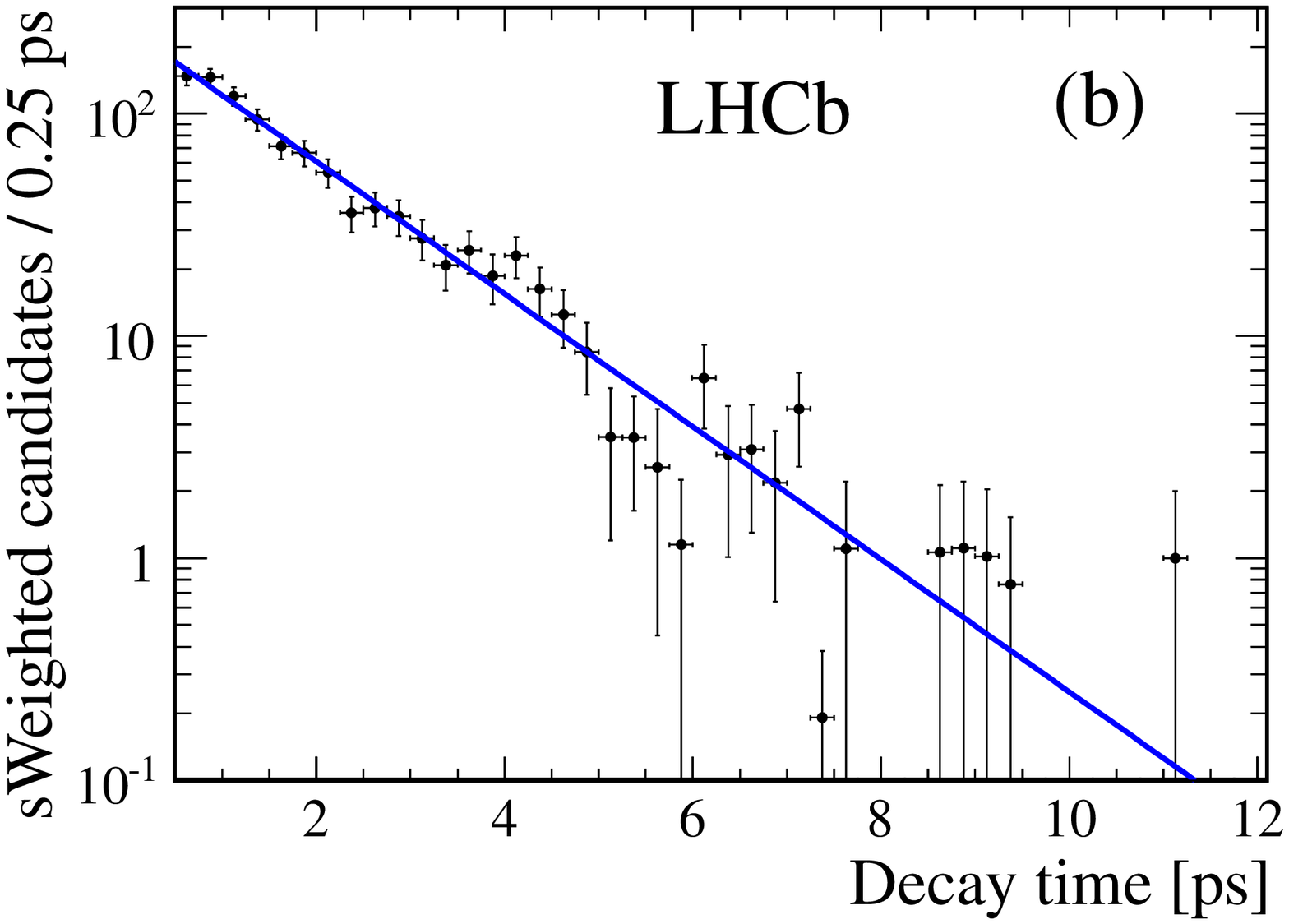,height=1.5in}
\caption{(a) Invariant mass spectrum for all selected $B^0_s \rightarrow K^+ K^-$ candidates. (b) Decay time distribution of $B^0_s \rightarrow K^+ K^-$ signal extracted using \textit{sWeights} and the fitted exponential function.}
\label{fig:KK}
\end{center}
\end{figure}
The analysis presented here~\cite{KK} uses the approach of selecting the $B^0_s$ mesons without biasing the decay time distribution, limiting the systematic uncertainties associated with correcting for any time-dependent acceptance effects.
This is achieved using neural networks in the software trigger and event selection, primarly using kinematic properties and particle identification requirements. The effective $B^0_s \rightarrow K^+ K^-$ lifetime is evaluated using an unbinned maximum log-likelihood fit. A fit to the invariant mass spectrum (Figure~\ref{fig:KK}a), which finds $997\pm34$ $B^0_s \rightarrow K^+ K^-$ decays, is performed to determine the \textit{sWeights}~\cite{sweight} that are used to isolate the $B^0_s \rightarrow K^+ K^-$ decay time distribution from the residual background. This distribution is fitted with a single exponential function and the effective $B^0_s \rightarrow K^+ K^-$ lifetime is found to be $\tau_{KK} = 1.455 \pm 0.046\, \mathrm{(stat.)} \pm 0.006\, \mathrm{(syst.)}$ $\mathrm{ps}$ in good agreement with the SM prediction of $1.40 \pm 0.02$ $\mathrm{ps}$. 
The main systematic contribution is coming from the bias in the lifetime, found after the reconstruction of the tracks in the final state, which is determined from simulated events.\\
\begin{figure}[ht]
\begin{center}
\epsfig{file=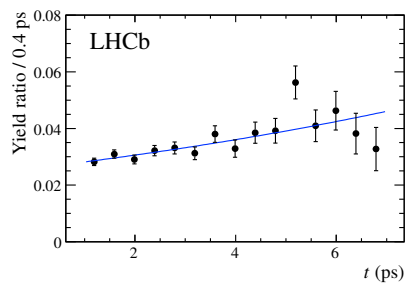,height=1.5in}
\caption{Decay time ratio between $\overline{B}^0_s \rightarrow J/\psi f_0(980)$ and $\overline{B}^0 \rightarrow J/\psi \overline{K}^{*0}(892)$ with the fit for $\Delta_{J/\psi f_0(980)}$ superimposed.}
\label{fig:f0}
\end{center}
\end{figure}
The $\overline{B}^0_s$ effective lifetime has been determined also in the channel $\overline{B}^0_s \rightarrow J/\psi f_0(980)$. The $f_0(980)$ selection results in a $\overline{B}^0_s \rightarrow J/\psi f_0(980)$ sample of $4040 \pm 75$ events, that is greater than $99.4\%$ $CP$-odd at $95\%$ confidence level~\cite{f0}. This implies that in the absence of $CP$ violation, the $J/\psi f_0(980)$ final state can be reached only by the decay of the heavy (H) $\overline{B}^0_s$ mass eigenstate. As the measured $CP$ violation in this final state is small~\cite{CPf0}, a measurement of the effective lifetime, $\tau_{J/\psi f_0(980)}$, can be translated into a measurement of the decay width, $\Gamma_H$.
In the analysis presented here~\cite{f0} the effective $\overline{B}^0_s \rightarrow J/\psi f_0(980)$ is measured relative to the well known measured $\overline{B}^0$ lifetime, in the decay mode $\overline{B}^0 \rightarrow J/\psi \overline{K}^{*0}(892)$, $\overline{K}^{*0}(892) \rightarrow K^{-} \pi^{+}$, from the variation of the ratio of the $B$ meson yields with decay time: $R(t) = R(0) e^{-t(1/\tau_{J/\psi f_0(980)} - 1/\tau_{J/\psi \overline{K}^{*0}(892)})} = R(0) e^{-t \Delta_{J/\psi f_0(980)}}$.
The same trigger and offline selection, using identical kinematic constraints, is applied to both channels, except for the hadron identification. The advantage of this technique is that in the ratio the decay time acceptance introduced by the trigger, reconstruction and selection requirements and the systematic uncertainties largely cancel. 
The decay time ratio between $\overline{B}^0_s \rightarrow J/\psi f_0(980)$ and $\overline{B}^0 \rightarrow J/\psi \overline{K}^{*0}(892)$ with the fitted probability density function (PDF) for $\Delta_{J/\psi f_0(980)}$ superimposed is shown in Figure~\ref{fig:f0}. Making use of a binned fit, the reciprocal lifetime difference is determined to be $\Delta_{J/\psi f_0(980)} = -0.070 \pm 0.014$ $\mathrm{ps}^{-1}$, where the uncertainty is statistical only. Taking $\tau_{J/\psi \overline{K}^{*0}(892)}$ to be the mean $\overline{B}^0$ lifetime $1.519 \pm 0.007$ $\mathrm{ps}$~\cite{pdg}, it is possible to determine $\tau_{J/\psi f_0(980)} = 1.700 \pm 0.040\, \mathrm{(stat.)} \pm 0.026 \,\mathrm{(syst.)}$ $\mathrm{ps}$. The main systematic contribution is due to non perfect cancelation between the acceptance corrections in the two channels.\\
The decay $B^0_s \rightarrow J/\psi \phi$ is considered the golden mode to measure $\phi_s$. It is a pseudo-scalar to vector-vector decay, so angular momentum conservation implies that the final state is an admixture of $CP$-even and $CP$-odd components. By performing a time-dependent angular analysis~\cite{CPf0}, it is possible to statistically disentangle the different $CP$ eigenstates by the differential decay rate for $B^0_s$ and $\overline{B}^0_s$ mesons produced as flavour eigenstates at $t=0$. Using $1.0$ $\mathrm{fb}^{-1}$ of data collected in $pp$ collisions at $\sqrt{s}=7$ TeV with the LHCb detector, the $CP$ violating phase $\phi_s$ as well as $\Delta \Gamma_s$ and $\Gamma_s$ are extracted by performing an unbinned maximum log-likelihood fit to the $B^0_s$ mass, decay time $t$, angular distributions and initial flavour tag of the selected $B^0_s \rightarrow J/\psi \phi$ events. The fitted decay time distribution is shown in Figure~\ref{fig:jpsiphi}a with the projections of the fitted PDF overlayed. \\
The 
results obtained from the fit are: $\Gamma_s = 0.6580 \pm 0.0054 \,\mathrm{(stat.)} \pm 0.0066 \,\mathrm{(syst.)}\,\mathrm{ps}^{-1}$ and $\Delta \Gamma_s = 0.116 \pm  0.018 \, \mathrm{(stat.)} \pm 0.006 \,\mathrm{(syst.)} \,\mathrm{ps}^{-1} \label{eq:deltagamma}$.\\
It must be noted that there exists another solution (solution II) for $\Delta \Gamma_s$ with the opposite sign, arising from the fact that the time-dependent differential decay rates are invariant under the transformation $(\phi_s,\Delta \Gamma_s) \leftrightarrow (\pi - \phi_s,-\Delta \Gamma_s)$, together with an appropriate transformation for the strong phases.
\begin{figure}[hb]
\begin{center}
\epsfig{file=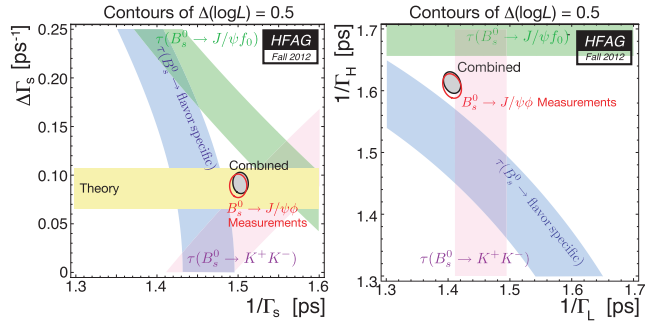,height=1.65in}
\caption{Contours of $\Delta(log(\mathcal{L})) = 0.5$ in the plane (1/$\Gamma_s$, $\Delta \Gamma_s$) on the left and in the plane (1/$\Gamma_L$, 1/$\Gamma_H$) on the right (see the text for the description).}
\label{fig:comb_final}
\end{center}
\end{figure}
The interference between the $K^+K^-$ S-wave and P-wave amplitudes in $B^0_s \rightarrow J/\psi K^+K^-$ decays, with the $K^+K^-$ pairs in the region around the $\phi(1020)$ resonance, is used to solve the ambiguity.
It can be shown that the physical solution is characterized by a difference between the $S$-wave and $P$-wave amplitudes which falls rapidly with increasing $m_{K^+K^-}$~\cite{sign}. Based on about $0.4\,\mathrm{pb}^{-1}$ of integrated luminosity the LHCb collaboration measured the phase difference reported in Figure~\ref{fig:jpsiphi}b: it is clear that only solution I is physical, and we conclude that the $B^0_s$ mass eigenstate that is almost $CP =+1$ is lighter and decays faster than the mass eigenstate that is almost $CP =-1$~\cite{sign}. This determines the sign of the decay width difference $\Delta \Gamma_s$ to be positive, as expected in the SM. \\
All results available publicly before the end of September 2012, including those described in these proceedings, have been included in the averages computed by the lifetime and oscillations sub-group of the Heavy Flavour Averaging Group (HFAG)~\cite{HFAG}. The two plots in Figure~\ref{fig:comb_final} show contours of $\Delta(log(\mathcal{L})) = 0.5$ ($39\%$ CL for the enclosed 2D regions, $68\%$ CL for the bands), in the plane (1/$\Gamma_s$, $\Delta \Gamma_s$) on the left and in the plane (1/$\Gamma_L$, 1/$\Gamma_H$) on the right. The average of all $B^0_s \rightarrow J/\psi \phi$ measurements is shown as the red contour, and the constraints given by the effective lifetime measurements of $\overline{B}^0_s \rightarrow J/\psi f_0(980)$, $\overline{B}^0_s \rightarrow K^+ K^-$ and $B^0_s$ to flavour-specific final states are shown as the green, purple and blue bands, respectively. The average taking all constraints into account is shown as the gray filled contour and the results are $\Delta \Gamma_s = 0.091 \pm 
0.011\,\mathrm{ps}^{-1}$ and $1/\Gamma_s = 1.503 \pm 0.010 \,\mathrm{ps}$. These values are in good agreement with the theory prediction $\Delta \Gamma_s = 0.087 \pm 0.021\, \mathrm{ps}^{-1}$, represented by the yellow band, which assumes no new physics in $B^0_s$ mixing~\cite{theory}.
\section{Conclusions}
The $B_s^0\,(\overline{B}_s^0)$ effective lifetime making use of the decay modes $B_s^0 \rightarrow K^+ K^-$ and $\overline{B}_s^0 \rightarrow J/\psi f_0(980)$ have been measured using $1.0$ $\mathrm{fb}^{-1}$ of data recorded by the LHCb detector. Using the same dataset also a measurement of $\Delta \Gamma_s$ and $\Gamma_s$ was made by a tagged time-dependent angular analysis of $B_s^0 \rightarrow J/\psi \phi$ decays and independently the sign of the decay width difference $\Delta \Gamma_s$ was established to be positive for the first time. All these measurements are the most precise measurements to date and are consistent with theoretical predictions~\cite{theory}.

\def\Discussion{
\setlength{\parskip}{0.3cm}\setlength{\parindent}{0.0cm}
     \bigskip\bigskip      {\Large {\bf Discussion}} \bigskip}
\def\speaker#1{{\bf #1:}\ }
\def\endDiscussion{}




 

\begin{thebibliography}{99}
{\small{
\bibitem{Lenz}
A. Lenz, Nucl. Phys. Proc. Suppl. 177-178:81-86 (2008).

\bibitem{Fleischer}
R. Fleischer and R. Knegjens, Eur. Phys. J. C71 1789 (2011).

\bibitem{LenzNierste}
A. Lenz and U. Nierste, JHEP 06 72 (2007).


\bibitem{LHCb}
LHCb collaboration, A. A. Alves Jr. \textit{et al.}, JINST 3 S08005 (2008).

\bibitem{KK}
LHCb collaboration, R. Aaij \textit{et al.}, Phys. Lett. B716 393-400 (2012).

\bibitem{sweight}
 Y. Xie, arXiv:0905.0724 [physics.data-an] (2009).

\bibitem{f0}
LHCb collaboration, R. Aaij \textit{et al.}, Phys. Rev. Lett. 109 152002 (2012).

\bibitem{CPf0}
LHCb collaboration, R. Aaij \textit{et al.}, LHCb-CONF-2012-002 (2012).

\bibitem{pdg}
Particle Data Group, J.Beringer \textit{et al.}, Phys. Rev. D86 010001 (2012).

\bibitem{sign}
LHCb collaboration, R. Aaij \textit{et al.}, Phys. Rev. Lett. 108 241801 (2012).

\bibitem{HFAG}
Y. Amhis et al. (Heavy Flavor Averaging Group), arXiv:1207.1158 [hep-ex] (2012).

\bibitem{theory}
A. Lenz and U. Nierste, TTP11-03, TUM-HEP-792/11 (2011).
}}
\end{thebibliography}
\end{document}